\documentclass[preprintnumbers,amsmath,amssymbm,preprint]{revtex4}
\usepackage{epsfig}
\usepackage{graphicx}

\begin{document}
\title{Dyonic black holes supporting nearly-black self-gravitating thin shells}
\author{Shahar Hod}
\address{The Ruppin Academic Center, Emeq Hefer 40250, Israel}
\date{\today}

\begin{abstract}
\ \ \ It has recently been revealed that dyonic black-hole spacetimes 
of a quasitopological non-linear electrodynamic field theory may be characterized by discrete radial 
regions with the property $df(r)/dr=0$, where $f(r)\equiv -g_{tt}(r)$, in which 
spherically symmetric massive {\it test} shells (Dyson shells with negligible self-gravity) 
can be supported in static equilibrium states.
In the present paper we prove that the dyonic spacetimes of the non-linear electrodynamic field theory 
may also be characterized by the presence of candidate {\it limiting} radii with the dimensionless 
property $d[r\cdot f(r)]/dr\to0^+$ in which massive {\it self-gravitating} thin shells that are 
on the verge of becoming black holes can be supported in static equilibrium states. 
It is proved that the discrete candidate limiting radii of these self-gravitating nearly-black Dyson shells are universal 
in the sense that they are independent of the masses of the central supporting dyonic compact objects. 
\end{abstract}
\bigskip
\maketitle


\section{Introduction}

The canonical family of electrically charged Reissner-Nordstr\"om black-hole spacetimes \cite{Chan} is 
known not to possess static equilibrium points at which massive test particles can remain at rest with respect 
to static asymptotic observers. 
This fact implies that test Dyson shells \cite{Dys} (massive test shells 
with negligible self-gravity) cannot be supported in static equilibrium states by charged 
Reissner-Nordstr\"om black holes \cite{Notesee,Isr,Hub,CL,Kij}. 

It has recently been revealed \cite{qua4} that a simple generalization of
electromagnetism, referred to as quasitopological non-linear electromagnetic 
field theory \cite{qua1} (see also \cite{qua2,qua3,Hoddyon1} and references therein) 
is characterized by the existence of dyonic black-hole spacetimes that may 
possess equilibrium radii $\{R_{\text{eq}}\}$, which are characterized by the gradient relation \cite{qua4}
\begin{equation}\label{Eq1}
{{df(r)}\over{dr}}=0\ \ \ \ \text{for}\ \ \ \ r=R_{\text{eq}}\  ,
\end{equation}
where $f(r)=-g_{tt}(r)$ [here $g_{tt}(r)$ is the radially-dependent metric function that 
characterizes the dyonic black-hole spacetimes, see Eq. (\ref{Eq5}) below], 
at which {\it test} particles can remain at rest a constant radial distance above the horizon of the central black 
hole . 

The observation made in \cite{qua4}, according to which the dyonic 
black holes of the non-linear electromagnetic field theory \cite{qua1} may be characterized 
by the property (\ref{Eq1}), implies that these non-vacuum curved spacetimes can support 
test Dyson shells \cite{Dys} in static equilibrium states. 

Motivated by the results presented in \cite{qua4}, in the present compact paper we shall use 
analytical techniques in order to prove that the dyonic black-hole spacetimes of the 
quasitopological non-linear electromagnetic field theory \cite{qua1} can also support {\it self-gravitating} 
nearly-black shells, massive thin shells that are on the verge of becoming black holes. 
[Note that, in the limiting configuration, it is the exterior geometry of the shell that develops a horizon, see 
Eq. (\ref{Eq21}) below.]  

In particular, below we shall explicitly prove that the critical spacetime relation \cite{Notegm}
\begin{equation}\label{Eq2}
{{d[r\cdot f(r)]}\over{dr}}\to0^+\ \ \ \ \text{for}\ \ \ \ r=R^{\text{c}}_{\text{eq}}\  ,
\end{equation}
which takes into account non-linear self-gravity effects, 
provides a necessary condition for the non-vacuum 
curved spacetimes to be able to support the nearly-black self-gravitating static thin shells. 

\section{Description of the system}

The quasitopological non-linear electrodynamic field theory is characterized by the action \cite{qua1,qua2,qua3,qua4,Noteunits}
\begin{equation}\label{Eq3}
S={{1}\over{16\pi}}\int{\sqrt{-g}d^4x\{R-\alpha_1 F^2-\alpha_2[(F^2)^2-2F^{(4)}]\}}\  .
\end{equation}
Here $F^2=F^{\mu\nu}F_{\mu\nu}$ and 
$F^{(4)}=F^{\mu}_{\nu}F^{\nu}_{\rho}F^{\rho}_{\sigma}F^{\sigma}_{\mu}$. 
The coupling parameter $\alpha_1$ in the action (\ref{Eq3}) is dimensionless. 
On the other hand, the non-trivial coupling parameter $\alpha_2$ in the composed 
action (\ref{Eq3}) of the non-linear electromagnetic field theory 
has the dimensions of length$^2$. 
As discussed in \cite{qua1}, the quasitopological electromagnetic field theory (\ref{Eq3}) 
may provide an interesting alternative to the standard Maxwell theory of electromagnetism. 

The dyonic black-hole solutions of the Einstein field equations coupled to the 
non-linear electromagnetic field theory (\ref{Eq3}) are characterized by a curved line element of the form
\cite{Whb,Shap,qua1,Notesch2}
\begin{equation}\label{Eq4}
ds^2=-f(r)dt^2 +{{1}\over{f(r)}}dr^2+r^2(d\theta^2 +\sin^2\theta d\phi^2)\  .
\end{equation}
The radially-dependent metric function in (\ref{Eq4}) is given by the functional expression \cite{qua1}
\begin{equation}\label{Eq5}
f(r;{\cal M},q,p,\alpha_1,\alpha_2)=
1-{{2{\cal M}}\over{r}}+{{\alpha_1 p^2}\over{r^2}}+{{q^2}\over{\alpha_1 r^2}}\cdot
{_2F_1}\Big({1\over4},1;{5\over4};-{{4\alpha_2p^2}\over{\alpha_1 r^4}}\Big)\  ,
\end{equation}
where ${\cal M}$ is the total (asymptotically measured) mass of the spacetime, 
$q$ is its electric charge, $p/\alpha_1$ is the 
magnetic charge \cite{Notenlg}, and ${_2F_1}(a,b;c;z)$ is the hypergeometric function \cite{Abram}. 
As discussed in \cite{qua1}, in the regime $\alpha_1>0$ with $\alpha_2>0$ the null, the weak, and the 
dominant energy conditions are respected by the dyonic black-hole 
solution (\ref{Eq5}), whereas the strong energy condition is violated. 

It is convenient to express the metric function (\ref{Eq5}) in the form \cite{Hodmn}
\begin{equation}\label{Eq6}
f(r)=1-{{2M(r)}\over{r}}\  ,
\end{equation}
where $M(r)$ is the gravitational mass contained within a sphere of radius $r$ with the 
asymptotic property 
\begin{equation}\label{Eq7}
M(r\to\infty)\to {\cal M}\  .
\end{equation}
If the spacetime contains a central black hole of horizon radius $r_{\text{H}}$ 
then its horizon mass is characterized by the simple property 
\begin{equation}\label{Eq8}
M(r=r_{\text{H}})={{r_{\text{H}}}\over{2}}\  .
\end{equation}

The radial equation of motion of a spherically symmetric self-gravitating thin shell of proper mass $m$ 
in the curved spacetime (\ref{Eq4}) is given by (see \cite{CL,Hub} and references therein) \cite{Notetau,Notedet}
\begin{equation}\label{Eq9}
{{m}\over{R}}=\sqrt{f_-(R)+\dot R^2}-\sqrt{f_+(R)+\dot R^2}\  ,
\end{equation}
where [see Eq. (\ref{Eq6})]
\begin{equation}\label{Eq10}
f_{\pm}(R)=1-{{2M_{\pm}(R)}\over{R}}\  .
\end{equation}
The mass difference across the shell is given by the relation
\begin{equation}\label{Eq11}
M_+(R)-M_-(R)=E(R)\  ,
\end{equation}
where $E(R;m)$ is the energy of the shell as measured by asymptotic observers \cite{Notemd}. 

\section{Static self-gravitating nearly-black shells in curved spacetimes} 

In the present section we shall determine the unique physical properties that a non-vacuum curved spacetime should have 
in order to be able to support self-gravitating nearly-black shells in static equilibrium states. 
To this end, we shall first determine the energy $E(R;m)$ of a spherically symmetric 
static shell of proper mass $m$ and radius $R$ which is on the verge of becoming a black hole. 

Substituting into Eq. (\ref{Eq9}) the relation 
\begin{equation}\label{Eq12}
\dot R\to0\
\end{equation}
for a static self-gravitating thin shell, one obtains the compact functional expression
\begin{equation}\label{Eq13}
{{m}\over{R}}=\sqrt{f_-(R)}-\sqrt{f_+(R)}\
\end{equation}
for the dimensionless proper-mass-to-radius ratio of the thin shell, where [see Eqs. (\ref{Eq10}) and (\ref{Eq11})]
\begin{equation}\label{Eq14}
f_+(R)=f_-(R)-{{2E(R)}\over{R}}\  .
\end{equation}

From Eqs. (\ref{Eq13}) and (\ref{Eq14}) one finds that, for a fixed proper mass $m$, the energy of the shell 
is given by the functional expression 
\begin{equation}\label{Eq15}
E(R;m)=m\sqrt{f_-(R)}-{{m^2}\over{2R}}\  .
\end{equation}
For the self-gravitating thin shell to be in a static equilibrium state, its energy 
as given by the radius-dependent relation (\ref{Eq15}), should be characterized by the extremum condition 
\begin{equation}\label{Eq16}
E'(R;m)=0\ \ \ \ \text{for}\ \ \ \ R=R_{\text{eq}}\  ,
\end{equation}
where a prime $'$ denotes a derivative with respect to the radial coordinate. 
Substituting Eq. (\ref{Eq15}) into Eq. (\ref{Eq16}), one obtains the dimensionless relation 
\begin{equation}\label{Eq17}
{{m}\over{R}}=-{{Rf'_-(R)}\over{\sqrt{f_-(R)}}}\ \ \ \ \text{for}\ \ \ \ R=R_{\text{eq}}\  .
\end{equation}

Taking cognizance of the junction condition (\ref{Eq13}) one finds from (\ref{Eq17}) the relation 
\begin{equation}\label{Eq18}
\sqrt{f_+(R)}={{f_-(R)+Rf'_-(R)}\over{\sqrt{f_-(R)}}}={{[Rf_-(R)]'}\over{\sqrt{f_-(R)}}}\ \ \ \ \text{for}\ \ \ \ R=R_{\text{eq}}\  ,
\end{equation}
which implies 
\begin{equation}\label{Eq19}
f_+(R)={{([Rf_-(R)]')^2}\over{f_-(R)}}\ \ \ \ \text{for}\ \ \ \ R=R_{\text{eq}}\  .
\end{equation}
The condition (\ref{Eq19}) characterizes genuine timelike equilibrium shells. 
From Eqs. (\ref{Eq17}), (\ref{Eq18}), and (\ref{Eq19}) one deduces that a static timelike shell with $f_+(R)>0$ requires
\begin{equation}\label{Eq20}
\big\{f_->0\ \ \ ; \ \ \ f'_-<0\ \ \ ; \ \ \ [Rf_-]'>0\big\}\ \ \ \ \text{for}\ \ \ \ R=R_{\text{eq}}\  .
\end{equation}

Our goal in this section is to determine the physical conditions that are required in order for 
the non-vacuum curved spacetime (\ref{Eq4}) to be able to support 
static self-gravitating {\it nearly-black} shells. 
These spherically symmetric matter configurations, which are on the verge of becoming black holes, 
are characterized by the limiting critical behavior 
\begin{equation}\label{Eq21}
f_+(R)\to0^+\
\end{equation}
of the metric function just outside the massive shell. 
Taking cognizance of Eqs. (\ref{Eq19}) and (\ref{Eq21}), one finds that static self-gravitating nearly-black shells, if they 
exist, are characterized by the limiting gradient relation 
\begin{equation}\label{Eq22}
\big[R\cdot f_-(R)\big]'\to0^+\ \ \ \ \text{for}\ \ \ \ R\to R^{\text{c}}_{\text{eq}}\  .
\end{equation}

It is worth emphasizing the fact that the analytically derived functional relation (\ref{Eq22}), 
which characterizes non-vacuum curved spacetimes that are able to support static 
self-gravitating nearly-black shells (thin shells that are on the verge of becoming black holes), is 
universal in the sense that it is independent of the proper masses of the supported shells. 
 
\section{The number of self-gravitating nearly-black shells that can be supported 
in the non-vacuum curved spacetimes}

In the present section we shall prove that, in general, the spherically symmetric non-vacuum 
black-hole spacetime (\ref{Eq4}) can support an even (or zero) number of self-gravitating nearly-black shells whose 
radii are determined by the dimensionless critical relation (\ref{Eq22}). 

To this end, we first point out that the boundary conditions \cite{Noteotex} 
\begin{equation}\label{Eq23}
\{f_-(r)=0 \ \ \ \text{and}\ \ \ f'_-(r)\geq0\}\ \ \ \ \text{for}\ \ \ \ r=r_{\text{H}}\  ,  
\end{equation}
which characterize the black-hole horizon, imply the inner boundary relation
\begin{equation}\label{Eq24}
\big[R\cdot f_-(R)\big]'\geq0\ \ \ \ \text{for}\ \ \ \ R\to r_{\text{H}}^+\  .
\end{equation}
In addition, using the relations
\begin{equation}\label{Eq25}
\{f_-(r)\to1 \ \ \ \text{and}\ \ \ rf'_-(r)\to0\}\ \ \ \ \text{for}\ \ \ \ r\to\infty\
\end{equation}
for asymptotically flat spacetimes, one finds the simple asymptotic behavior
\begin{equation}\label{Eq26}
\big[R\cdot f_-(R)\big]'\to1\ \ \ \ \text{for}\ \ \ \ R\to\infty\  .
\end{equation}

The characteristic functional behaviors (\ref{Eq24}) and (\ref{Eq26}) imply that the 
dimensionless function $[R\cdot f_-(R)]'$ 
has in general \cite{Noteing1} an even (or zero) number of roots in the exterior region of the black-hole spacetime. 
This fact implies that the non-vacuum curved spacetime (\ref{Eq4}) can support 
an even (or zero) number of self-gravitating nearly-black shells. 
As recently proved in \cite{qua4}, this is also the case 
for the functional condition $[f_-(R)]'=0$ [see Eq. (\ref{Eq1})] that 
characterizes spacetime regions in which test shells (shells with negligible self-gravity) 
can be supported in static equilibrium states.

It is important to note that the even number of nearly-black shells characterizes 
simple exterior roots of the function $[Rf_-]'$ [this is because the function $[Rf_-]'$ has the same positive 
sign at the two boundaries, see Eqs. (\ref{Eq24}) and (\ref{Eq26})]. 
It should be realized, however, that if degenerate roots exist then the 
number of distinct limiting roots may not be even.

\section{Dyonic black holes supporting self-gravitating nearly-black thin shells}

In the present section we shall use the analytically derived results of the previous sections in order to 
analyze the physical and mathematical properties of self-gravitating thin shells that are on the 
verge of becoming black holes in the dyonic curved spacetime (\ref{Eq5}) \cite{Noteclar}. 

Substituting the metric function (\ref{Eq5}) of the dyonic black holes 
into the analytically derived functional relation (\ref{Eq22}) 
and using the characteristic gradient relation \cite{Abram}
\begin{equation}\label{Eq27}
{{d\Big[{_2F_1}\Big({1\over4},1;{5\over4};-{{4\alpha_2p^2}\over{\alpha_1 r^4}}\Big)\Big]}\over{dr}}=
{{{_2F_1}\Big({1\over4},1;{5\over4};-{{4\alpha_2p^2}\over{\alpha_1 r^4}}\Big)}\over{r}}-
{{r^3}\over{r^4+{{4\alpha_2p^2}\over{\alpha_1}}}}\
\end{equation}
of the hypergeometric function, one finds that the polynomial equation
\begin{equation}\label{Eq28}
{\cal F}(R;q,p,\alpha_1,\alpha_2)=0\ \ \ \ \text{for}\ \ \ \ R=R^{\text{c}}_{\text{eq}}\
\end{equation}
with
\begin{equation}\label{Eq29}
{\cal F}(R;q,p,\alpha_1,\alpha_2)\equiv\alpha_1\cdot R^6-(\alpha_1^2p^2+q^2)\cdot R^4+4\alpha_2p^2\cdot R^2-4\alpha_1\alpha_2p^4\
\end{equation}
determines the candidate limiting radii $\{R^{\text{c}}_{\text{eq}}\}$ of static self-gravitating 
nearly-black shells. 

Inspection of Eqs. (\ref{Eq28}) and (\ref{Eq29}) reveals the fact that 
the critical radii $\{R^{\text{c}}_{\text{eq}}(q,p,\alpha_1,\alpha_2)\}$, which 
characterize self-gravitating nearly-black thin shells that are located around dyonic black holes, 
do not depend on the asymptotically measured masses ${\cal M}$ of the supporting curved 
spacetimes \cite{Notennn}. 

As we shall now prove explicitly, the cubic equation (\ref{Eq28}) can be used to derive necessary conditions 
for the existence of static self-gravitating nearly-black shells in the dyonic black-hole spacetime (\ref{Eq5}). 
To this end, we first note that the function ${\cal F}(R;q,p,\alpha_1,\alpha_2)$ is characterized by the 
properties 
\begin{equation}\label{Eq30}
{\cal F}(R\to0;q,p,\alpha_1,\alpha_2)<0\ \ \ \ \text{and}\ \ \ \ {\cal F}(R\to\infty;q,p,\alpha_1,\alpha_2)\to\infty\  ,
\end{equation}
which imply that the function ${\cal F}(R;q,p,\alpha_1,\alpha_2)$ has at least one positive root. 

In order for the function ${\cal F}(R;q,p,\alpha_1,\alpha_2)$ to have three positive roots 
[as we have proved in the previous section, the spherically symmetric non-vacuum black-hole spacetime (\ref{Eq4}) 
generally supports an even (or zero) number of self-gravitating nearly-black shells in the external 
region $R>r_{\text{H}}$] it must have a local minimum point $R=R^{+}_{\text{ext}}$ with the property 
\begin{equation}\label{Eq31}
{\cal F}(R^{+}_{\text{ext}};q,p,\alpha_1,\alpha_2)<0\
\end{equation}
and a local maximum point $R=R^{-}_{\text{ext}}$ with the property 
\begin{equation}\label{Eq32}
{\cal F}(R^{-}_{\text{ext}};q,p,\alpha_1,\alpha_2)>0\  .
\end{equation}

The extremum points of the function ${\cal F}(R;q,p,\alpha_1,\alpha_2)$ are located at [see Eq. (\ref{Eq29})]
\begin{equation}\label{Eq33}
R^{\pm}_{\text{ext}}(q,p,\alpha_1,\alpha_2)=\sqrt{
{{\alpha_1^2p^2+q^2\pm\sqrt{(\alpha_1^2p^2+q^2)^2-12\alpha_1\alpha_2p^2}}\over{3\alpha_1}}}\  .
\end{equation}
The asymptotic functional behavior ${\cal F}(R\to\infty;q,p,\alpha_1,\alpha_2)\to\infty$ [see Eq. (\ref{Eq30})] 
implies that the smaller extremum radius $R^{-}_{\text{ext}}$ corresponds to a local maximum point 
of the function ${\cal F}(R;q,p,\alpha_1,\alpha_2)$ whereas 
the larger extremum radius $R^{+}_{\text{ext}}$ (with $R^{-}_{\text{ext}}\leq R^{+}_{\text{ext}}$) 
corresponds to a local minimum point of the function ${\cal F}(R;q,p,\alpha_1,\alpha_2)$ \cite{Noteinf}. 

In order to analyze the complete classification of the polynomial equation (\ref{Eq28}) 
it is convenient to define the dimensionless variables 
\begin{equation}\label{Eq34}
y\equiv {{R^2}\over{\alpha_1 p^2}}\ \ \ \ ; \ \ \ \ \beta\equiv {{\alpha_2}\over{\alpha^3_1 p^2}}\ \ \ \ ; \ \ \ \ 
\gamma\equiv {{q}\over{\alpha_1 p}}\  ,
\end{equation}
in terms of which the polynomial equation (\ref{Eq28}) can be written in the form
\begin{equation}\label{Eq35}
P(y)=y^3-(1+\gamma^2)\cdot y^2+4\beta\cdot y-4\beta=0\  .
\end{equation}
From the cubic equation (\ref{Eq35}) one finds the following classification: 
\begin{itemize}
\item In the regime $\gamma < 2\sqrt{2}$ the polynomial equation (\ref{Eq35}) has only one positive real root.

\item For $\gamma = 2\sqrt{2}$, the special value $\beta = 27/4$ gives the positive triple root $y = 3$.

\item In the regime $\gamma > 2\sqrt{2}$ the polynomial equation (\ref{Eq35}) has three distinct positive roots 
if and only if $\beta_{-}<\beta<\beta_{+}$ where 
$\beta_{\pm}=[{{\gamma^4+20\gamma^2-8\pm{\gamma(\gamma^2-8)^{3/2}}}]/{32}}$, 
while $\beta=\beta_{\pm}$ is characterized by one simple and one double positive root \cite{Notedegb}, 
and outside this interval the polynomial (\ref{Eq35}) has only one positive real root.
\end{itemize}

It is important to emphasize the fact that the candidate limiting radii, which are determined from 
the limiting polynomial (\ref{Eq35}), are independent of ${\cal M}$, but their existence
as exterior configurations depends on the mass-dependent condition \cite{Notennn}
\begin{equation}\label{Eq36}
R^{\text{c}}_{\text{eq}}>r_{\text{H}}({\cal M})\  .
\end{equation}

\section{Dyonic black holes supporting self-gravitating nearly-black thin shells: 
A numerically constructed example}

As a concrete example for a non-vacuum spacetime that can support self-gravitating nearly-black 
static shells, we shall now consider the dyonic curved spacetime (\ref{Eq5}) 
with the dimensionless physical parameters 
\begin{equation}\label{Eq37}
\alpha_1=1\ \ \ ; \ \ \ {{\alpha_2}\over{{\cal M}^2}}=4\ \ \ ; \ \ \ {{q}\over{{\cal M}}}=1.06
\ \ \ ; \ \ \ {{p}\over{{\cal M}}}=0.13\  .
\end{equation}

Taking cognizance of Eq. (\ref{Eq5}) with the characteristic horizon condition
\begin{equation}\label{Eq38}
f(r=r_{\text{H}};{\cal M},q,p,\alpha_1,\alpha_2)=0\  ,
\end{equation}
one finds that the horizon radius of the 
dyonic black hole (\ref{Eq37}) is characterized by the dimensionless relation \cite{Noteih}
\begin{equation}\label{Eq39}
{{r_{\text{H}}}\over{{\cal M}}}\simeq0.1896357\  .
\end{equation}
In addition, from Eqs. (\ref{Eq28}) and (\ref{Eq29}) one finds that the dyonic black-hole spacetime (\ref{Eq5}) with the 
physical parameters (\ref{Eq37}) can support 
two self-gravitating nearly-black thin shells which are characterized by the dimensionless critical 
{\it limiting} radii \cite{Notetl,Notelim,Notenumdtl,Notemnmx,Notenbnb} 
\begin{equation}\label{Eq40}
{{R^{\text{c}}_{\text{eq,1}}}\over{{r_{\text{H}}}}}\simeq2.918111\ \ \ \ \text{and}
\ \ \ \ {{R^{\text{c}}_{\text{eq,2}}}\over{{r_{\text{H}}}}}\simeq4.763444\  .
\end{equation}
As a consistency check we note that the central supporting dyonic black hole (\ref{Eq5}) with the 
physical parameters (\ref{Eq37}) is characterized by the relations $\gamma > 2\sqrt{2}$ and 
$\beta_{-}<\beta<\beta_{+}$.

\section{Summary and Discussion}

It has recently been revealed \cite{qua4} that dyonic black holes of the 
quasitopological electromagnetic field theory \cite{qua1} (see also \cite{qua2,qua3,Hoddyon1}) may 
possess special equilibrium radii $\{R_{\text{eq}}\}$ at which test Dyson shells 
(massive test shells with negligible self-gravity) can be supported in static equilibrium states. 

Motivated by the observation made in \cite{qua4}, in the present compact paper we have explored 
the physical and mathematical properties of {\it self-gravitating} nearly-black shells, 
massive Dyson shells which are on the verge of becoming black holes. 
In particular, we have determined, using analytical techniques, the physical conditions that are required for the non-vacuum 
dyonic spacetimes to be able to support self-gravitating nearly-black shells in static equilibrium states. 

The main analytical results derived in this paper and their physical
implications are as follows:

(1) Using the Israel junction condition, we have explicitly proved that self-gravitating nearly-black 
shells can be supported in a spherically symmetric non-vacuum 
spacetime of the form (\ref{Eq4}). In particular, spacetime regions that contain self-gravitating nearly-black shells, if they 
exist, are characterized by the {\it limiting} gradient relation 
\begin{equation}\label{Eq41}
{{d[R\cdot f_-(R)\big]}\over{dR}}\to0^+\ \ \ \ \text{for}\ \ \ \ R\to R^{\text{c}}_{\text{eq}}\  ,
\end{equation}
where $R^{\text{c}}_{\text{eq}}$ is the {\it candidate limiting radius} of the supported thin shell.

(2) It is worth emphasizing that the analytically derived critical spacetime relation (\ref{Eq41}) should be 
regarded as a modification, due to non-linear self-gravity effects, 
of the gradient condition (\ref{Eq1}) which characterizes 
the radii of test shells with negligible self-gravity \cite{Notegm}. 
 
(3) We have proved that the critical candidate limiting radii $\{R^{\text{c}}_{\text{eq}}(q,p,\alpha_1,\alpha_2)\}$ of the nearly-black thin shells in the dyonic black-hole spacetime (\ref{Eq5}) 
are independent of the asymptotically measured masses ${\cal M}$ of the supporting spacetimes.  
 
(4) It has been proved that a non-vacuum dyonic black-hole spacetime can generally have 
an even (or zero) number \cite{Noteing1} of candidate limiting radii 
that satisfy the critical functional relation (\ref{Eq41}) with the property $R^{\text{c}}_{\text{eq}}>r_{\text{H}}$. 
Thus, our results reveal the fact that dyonic black holes can generally support an even (or zero) number of self-gravitating nearly-black shells. 

It should be emphasized that the even number of nearly-black shells characterizes 
simple exterior roots (because the function $[Rf_-]'$ has the same positive sign 
at the two endpoints). If degenerate roots exist then the number of distinct limiting 
roots may not be even.
 
(5) We have proved that non-vacuum dyonic black-hole spacetimes that can support non-degenerate 
self-gravitating nearly-black shells 
in static equilibrium states are characterized by the compact dimensionless necessary condition [see 
Eqs. (\ref{Eq34}) and (\ref{Eq35})] \cite{Notedgn}
\begin{equation}\label{Eq42}
{{\gamma^4+20\gamma^2-8-{\gamma(\gamma^2-8)^{3/2}}}\over{32}}
<{{\alpha_2}\over{\alpha_1^3p^2}}<
{{\gamma^4+20\gamma^2-8+{\gamma(\gamma^2-8)^{3/2}}}\over{32}}
\ \ \ \ \text{for}\ \ \ \gamma>2\sqrt{2}\  ,
\end{equation}
where 
\begin{equation}\label{Eq43}
\gamma\equiv {{q}\over{\alpha_1 p}}\  .
\end{equation}

\bigskip
\noindent
{\bf ACKNOWLEDGMENTS}
\bigskip

This research is supported by the Carmel Science Foundation. 
I would like to thank the anonymous referee for valuable comments and suggestions. I also thank
Yael Oren, Arbel M. Ongo, Ayelet B. Lata, and Alona B. Tea for 
stimulating discussions.


\begin{thebibliography}{99}
\bibitem{Notegm} Note that the gradient relation (\ref{Eq2}) 
together with the property $f(r)>0$, which characterizes the exterior region of the curved spacetime, imply the 
inequality $df(r)/dr<0$ at the critical equilibrium radius $r=R^{\text{c}}_{\text{eq}}$ of a 
nearly-black self-gravitating thin shell.
\bibitem{Noteunits} We use natural units in which $G=c=1$.
\bibitem{Notesch2} We use the familiar Schwarzschild coordinates $\{t,r,\theta,\phi\}$ in the 
line element (\ref{Eq4}) of the spherically symmetric curved spacetime.
\bibitem{Notenlg} We shall assume the relations $q>0$ and $p>0$ without loss of generality.
\bibitem{Notetau} Here $\dot R\equiv dR/d\tau$, where $\tau$ is the proper time along the 
worldline of the thin massive shell \cite{Hub}.
\bibitem{Notedet} We consider dust shells with vanishing tangential pressures 
which are electrically and magnetically neutral with 
no surface charges and currents. 
The proper mass of the shell is constant during its radial motion. 
The physical parameters $q$ and $p$ are identical on both sides of the thin shell.
\bibitem{Notemd} It is worth noting that the 
mass-function difference is constant if the electromagnetic charges and couplings agree 
on the two sides of the shell.
\bibitem{Noteotex} Note that the gradient relation $f'_-(r=r_{\text{H}})\geq0$ characterizes the 
outermost horizon of the central black hole, where the equality sign 
corresponds to an extremal (zero-temperature) black hole.
\bibitem{Noteing1} Note that the radial function $[R\cdot f_-(R)]'$ 
may possess degenerate roots. The even-root argument applies to simple roots; degenerate
roots can change the number of distinct radial roots.
\bibitem{Noteclar} It is worth emphasizing again that, in the limiting configuration, 
it is the exterior geometry of the shell that develops a horizon [see Eq. (\ref{Eq21})].
\bibitem{Notennn} It is important to emphasize that the candidate critical radii, which 
are determined by Eqs. (\ref{Eq28}) and (\ref{Eq29}), are mass-independent, 
whereas the existence and number of exterior shells are not universal in ${\cal M}$. 
In particular, a candidate shell radius must also satisfy the inequality $R^{\text{c}}_{\text{eq}}>r_{\text{H}}({\cal M})$, 
where the black-hole horizon radius $r_{\text{H}}({\cal M})$ depends on the mass parameter.
\bibitem{Noteinf} Note that the function ${\cal F}(R;q,p,\alpha_1,\alpha_2)$ 
has an inflection point at $R^{-}_{\text{ext}}=R^{+}_{\text{ext}}=(4\alpha_2p^2/3\alpha_1)^{1/4}$ 
for $\alpha_2=(\alpha_1^2p^2+q^2)^2/12\alpha_1p^2$.
\bibitem{Notedegb} Solving $P(y)=P'(y)=0$ [see Eq. (\ref{Eq35})] one finds, for $\gamma>2\sqrt{2}$, that 
the degenerate boundaries $\beta=\beta_{\pm}$ of the system correspond to the case in which 
the polynomial equation (\ref{Eq35}) has a double root 
at $y_{\text{d}_{\pm}}={{\gamma^2+4\pm\gamma\sqrt{\gamma^2-8}}\over{4}}$ and a single root at $y_{\text{s}_{\pm}}={{\gamma^2-2{\mp}\gamma\sqrt{\gamma^2-8}}\over{2}}$.
\bibitem{Noteih} From Eq. (\ref{Eq38}) one finds that the dyonic black-hole spacetime (\ref{Eq5}) 
with the dimensionless physical parameters (\ref{Eq37}) has an inner horizon which 
is characterized by the dimensionless relation $r_{\text{in}}/{\cal M}\simeq0.0977738$.
\bibitem{Notelim} It should be emphasized that the two radii (\ref{Eq40}) are roots of the limiting 
equation ${\cal F}(R;q,p,\alpha_1,\alpha_2)\to0$ [see Eqs. (\ref{Eq22}) and (\ref{Eq28})].
\bibitem{Notetl} Note that the third positive solution of the critical equation (\ref{Eq28}) 
is characterized by the dimensionless relation ${{R^{\text{c}}_{\text{eq}}}/{{r_{\text{H}}}}}\simeq0.713119$ 
and it is therefore located inside the black hole.
\bibitem{Notenumdtl} Note that one obtains from Eqs. (\ref{Eq13}), (\ref{Eq14}), and (\ref{Eq21}) the dimensionless 
relations ${{m}/{R}}=\sqrt{f_-(R)}$ and ${{E(R)}/{R}}={{f_-(R)}/{2}}$ for the static 
nearly-black self-gravitating thin shells. Taking cognizance of Eqs. (\ref{Eq5}), (\ref{Eq37}), (\ref{Eq39}), 
and (\ref{Eq40}) one finds the dimensionless 
relations $\{f_-=0.196246$, $m/{\cal M}=0.245145$, $E/{\cal M}=0.054299\}$ if the static 
shell is located at the inner radius $R=R^{\text{c}}_{\text{eq,1}}$,  
and $\{f_-=0.091599$, $m/{\cal M}=0.273392$, $E/{\cal M}=0.041372\}$ if the static 
shell is located at the outer radius $R=R^{\text{c}}_{\text{eq,2}}$.
\bibitem{Notemnmx} The stability properties of the nearly-black shells can be inferred from the 
signs of $E''(R;m)$ evaluated separately at the critical candidate limiting radii $R^{\text{c}}_{\text{eq}}$. 
In particular, from Eq. (\ref{Eq15}) one finds the relation $E''(R^{\text{c}}_{\text{eq}};m)=R^{\text{c}}_{\text{eq}}f''_-(R^{\text{c}}_{\text{eq}})/2-5f_-(R^{\text{c}}_{\text{eq}})/(4R^{\text{c}}_{\text{eq}})$ 
[we have used here the relation $m=R^{\text{c}}_{\text{eq}}\sqrt{f_-(R^{\text{c}}_{\text{eq}})}$, see Eqs. (\ref{Eq13}) and (\ref{Eq21}). In addition, we have used here the relation 
$f'_-(R^{\text{c}}_{\text{eq}})=-f_-(R^{\text{c}}_{\text{eq}})/R^{\text{c}}_{\text{eq}}$, see Eq. (\ref{Eq22})], 
which yields ${\cal M}E''(R=R^{\text{c}}_{\text{eq,1}})\simeq-0.81699<0$ 
and ${\cal M}E''(R=R^{\text{c}}_{\text{eq,2}})\simeq0.45546>0$. 
The radius-dependent energy of the shell therefore 
has a local maximum at the inner equilibrium radius $R^{\text{c}}_{\text{eq,1}}$, which implies that this limiting 
shell is radially unstable. 
On the other hand, the radius-dependent energy of the shell 
has a local minimum at the outer equilibrium radius $R^{\text{c}}_{\text{eq,2}}$, which implies that this limiting 
shell is radially stable.
\bibitem{Notedgn} It is worth noting that the condition ${\cal F}(R^+_{\text{ext}};q,p,\alpha_1,\alpha_2)<0$ yields 
the inequality $\alpha_2<27\alpha^3_1p^2/4$ for the degenerate case $q=2\sqrt{2}\alpha_1 p$.
\end{thebibliography}
\end{document}